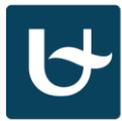

**Postgraduate course Energy & Climate**

**Academic year 2013-2014**

**Investigation on Demand Side Management Techniques in the Smart Grid using Game Theory and ICT Concepts**

**Bart Smets**
**bart.smets@gmail.com**

**Tutor: A. Verbruggen**

**August 2014**

# Contents

# Abstract

This paper investigates how concepts from game theory and ICT can contribute to solve challenges in demand side management, an important concept in the upcoming smart grid. Demand side management is about modifying the energy load distribution on the demand side, for example in order to reduce peaks in energy usage. This can be done by shifting energy demands where possible.

We start with describing a number of smart grid concepts and assumptions (smart meters, pricing, appliance scheduling) and explain the advantages demand side management has. After the introduction of game theoretic concepts, it becomes possible to mathematically describe the demand side management problem.

Next step is to solve the mathematical formulation, and show how complex demand side management becomes if the number of energy users increases. By means of distributed ICT algorithms however, it is possible to still find a solution.

Based on existing literature, different algorithms are studied. Though results in literature looked promising, several conclusions on convergence of the algorithm in general, and convergence towards the most optimal results in particular are challenged.

# 1 Introduction

A sustainable electrification over the coming decades is a widely suggested solution to ban fossil fuels. For example in transport the conventional fossil fuel cars could be replaced by electric vehicles and for heating purposes electrical heat pumps could be used [1].

However, this increase in electricity usage creates challenges for the electricity grid. One of these challenges is that the grid has to be dimensioned to deliver the peak demand in electricity. If electricity usage rises, the peaks in electricity usage could rise as well. An important focus of demand side management is to diminish these peaks by shifting electricity usage away from peak periods wherever possible.

In the past decades, ICT has gone through a revolution of changes. It is now possible to transfer huge amounts of data at the speed of light. Also, computer processing power has increased, while prices decreased, making it possible to make complex calculations at very low cost.

Game theory is a framework of mathematical models that helps studying strategic decision making. Since the 1950s, it is used in economics, in computer science and in other disciplines. The problems of demand side management can often be considered as games. These games are then addressed with known approaches from game theory.

The main focus of this paper is to investigate how the current possibilities in ICT, together with game theoretic concepts, could help solve upcoming demand side management challenges in smart grids. Our investigation is mainly on domestic demand side management.

The remainder of this paper is organised as follows. In Section 2 the most important concepts in smart grids are described, together with assumptions that will be used later. In Section 3 the challenges on demand side management are introduced. In Section 4 concepts of game theory are explained. In Section 5 these concepts are applied to the challenges in demand side management, creating a mathematical formulation of the problem. In Section 6 several algorithms that aim to solve this mathematical formulation are investigated and critically assessed. In Section 7 suggestions for future research are given. Section 8 contains the conclusion of this paper.

## 2  Smart Grid Concepts and Assumptions

### 2.1  Smart meters

We assume that in the near future buildings will be equipped with a smart meter and smart appliances. The smart meter communicates with these smart appliances. The smart meter is able to receive information from the smart appliances about energy consumption, the usage pattern etc. It is also able to schedule these appliances. If the building would deliver energy (e.g. from solar panels on the roof, or from the battery of an electric car) the smart meter communicates with these energy sources.

The smart meter also communicates with the grid. It receives current price information and price forecasts from the grid, and can send expected energy usage and energy production profiles to the grid. This way, the smart meter is in fact an intermediate between the grid and the building.

### 2.2  Electricity pricing

Today domestic users in general have fixed electricity tariffs, often split into one peak and one off-peak price. For example, electricity used at 19h in the evening on a cloudy cold Tuesday in February, could be charged the same price as electricity used at 9h in the morning on a sunny Wednesday in May. While for the electricity supplier, the price to produce this energy will be very different.

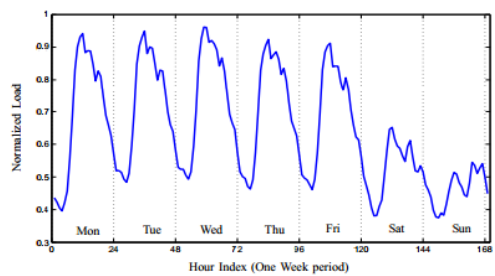

**Figure 1: Normalized weekly energy load on the Belgian grid [2]**

Figure 1 presents a typical weekly energy load profile on the Belgian grid [2]. More than twice as much energy is used at peak level, than at bottom level. A fixed part of the energy (the lower end of the graph in figure 1) is delivered by base load plants, like nuclear, coal, hydro, biomass and geothermal installations. These plants produce continuous power, at low cost. The peaks in energy usage are delivered by peak power plants (for example gas turbines). Compared to the amount of energy they produce, peak power plants are expensive. The flatter the energy usage curve, the more energy could be provided by base load plants, and the cheaper energy would be for everybody. Also, the grid (all its cables, stations, substations, transformers, etc.) is sized to be able to handle this peak value. The lower the peak value, the cheaper the grid.

### 2.2.1 Dynamic Pricing

Current electricity tariffs remain static throughout weeks, months, or even (for some contracts) years. Apart from a very basic day/night difference there is no monetary incentive for residential clients to shift their demand for example from the peak at 9h to the afternoon dip of 15h.

In a smart grid, it will be possible to introduce a far more dynamic way of pricing. Energy prices could change by the hour, and smart meters could be informed on current and future prices. The technology of the smart grid makes dynamic pricing possible. Because the energy price can be an important incentive to lower the expensive peak energy usage, it can be expected that dynamic pricing will be introduced in the future [1].

### 2.2.2 Price function

If all users decide to use a lot of energy at 9h, the electricity provider will have to use expensive peak plants to be able to provide this electricity. This expensive production cost will have its impact on the electricity price at 9h. A linear increase of the electricity used (for example +5%), will trigger a more than linear increase of the electricity unit price (for example +25%).

When a large part of the energy supply comes from intermittent renewable energy sources, the pricing function will be more complex. Wind and solar energy cannot be scheduled. It can however be predicted a few days in advance, based on weather models. At times when a lot of renewable energy is available, it will be good to have a larger energy load, so the price per unit will in that case not necessarily increase if the total load increases.

## 2.3 Appliance scheduling

Not every appliance has to run immediately when requested. For example: When using a dishwasher, the goal is to have clean dishes by tomorrow morning, not to start washing immediately after a user pressed start. A special case is an electric car, or Plug-in Electrical Vehicle (PEV). The goal could be to have a charged battery by morning; it does not matter when exactly the battery is charged. We could even go one step further: if the battery is charged and energy prices go up, it could be interesting to use the energy of the battery for in-house appliances, or even to sell this energy to the grid.

On the other hand, when turning on the television, the goal is to watch television immediately. Television, radio, a hairdryer and light are examples of appliances not suitable for shifting.

End users have monetary motivations, they want to pay as less as possible for their electricity, with a maximum of comfort. They want to be able to watch television and turn on lights immediately. For shiftable appliances, they might want to shift the energy usage, but at a maximum of comfort. Nobody wants to sit in front of a price feed to find out when is the best moment to start a dishwasher. This should all happen automated. The smart grid and smart appliances make this possible.

One can also expect that some users have additional motivations. For example, they want to use as much clean energy as possible. This can be included in the user's contract. It could also be achieved in a more global way for all users, by making renewable energy cheaper and non sustainable energy more expensive.

# 3   Challenges in Demand Side Management

In the past, too often the focus of energy companies was to match energy supply with the demand. Demand side management reverses this principle, and aims at modifying the usage pattern on the demand side, in order to match it with the energy supply. This is often done by financial methods: charging more for energy in peak periods, will lower the peak demand.

Some users will decide to turn of certain appliances (light, air conditioning etc.), and this will lower peak demand. Other users will decide to wait with certain tasks until after peak time. The latter is the topic of this paper. As we saw from the previous section, this decision can be taken by a smart meter.

However, with real time pricing and immediate response from the smart meter, this could cause an instable situation! Imagine the following scenario: At a given moment in time, energy is cheap. Owners of a smart meter now have a big advantage. The smart meter will communicate with the grid, and become aware of the cheap prices. The smart meter can then decide to use this cheap energy for one or more appliances (for example: start charging an electric car, or start the dishwasher), on the fly, without any human intervention.

The above scenario sounds tempting, but holds big risks: If a large percentage of households is equipped with a smart meter, this could mean that suddenly a large number of appliances practically simultaneously start to consume electricity.

From section 2.2.2, we expect prices to rise again. Smart meters will decide to take actions on this new situation, trying to use as little energy as possible. This can be done by postponing the charging of an electric car, postponing the dishwasher etc., or even by using the battery of an electric car as an energy source. This could go on for a while. Hopefully, this chain of reactions will end in a balanced situation, an equilibrium where a price level is reached where smart meters decide to make no changes to the current in house energy usage. A similarity can be seen with any other good of economic value. For example everybody knows strawberries are cheaper in summer than in winter, because of high supply in summer. As a consequence, more people buy strawberries in summer than in winter. Based on the sticker price, customers will decide to buy or not to buy strawberries.

From a technical point of view however, there is a big difference with most other consumer goods: time. The order of magnitude in time when considering consumer goods in a supermarket is days. It will take a few days before all customers are aware of the cheap price of strawberries. The order of magnitude of time when considering ICT concepts is microseconds. Smart meters could react in a fraction of a second, causing a large number of appliances to simultaneously start or stop. This is not desired; a sudden change in the electricity consumption possibly causes instabilities and could even end in a blackout.

At first sight, the new constellation made it easier for end users to enjoy cheap energy prices, but the above scenario shows that the introduction of smart meters will inevitably introduce a whole new way of pricing and consuming electricity. In the following two sections, this is translated into a mathematical model.

# 4 Game Theory Concepts

The scenario from the previous paragraph can be described and simulated with concepts from Game Theory.

The smart meters deciding whether to buy energy at a certain moment or not, taking into account the expectations of the house owner (for example the car has to be fully charged by morning), are in fact playing a *Noncooperative Game*. A smart meter decision is based on the energy price, while the energy price is dependent on what the other smart meters decide.

In [3] and [4] important concepts from game theory are introduced. This serves as a base for this section.

## 4.1 Definition 1: a Game

A **game** is defined as a situation involving three components:

1. A set of players $N$
2. For every player, a set of possible strategies $(S_i)_{i \in N}$
3. For every player, a utility function $(u_i)_{i \in N}$

The set of players is in our case the set of households.

The set of possible strategies of a player $i$ is the collection of all possible options the player has. If the game was to guess a number between 1 and 10, the set of possible strategies would be any number between 1 and 10.

The utility function represents the motivations of players. It assigns a value to every possible outcome of the game. It is often expressed as an amount of money, also in this paper. However, please note that this function can very well be something else. Anything that can be quantified can serve as a utility function. Fictitious example: a user quantifies his car to be loaded before midnight as 5 points. This function depends on the strategy chosen by the player, but also on the strategies chosen by all other players. It is denoted as $u_i(s_i, \vec{s}_{-i})$ where $\vec{s}_{-i}$ is the vector of strategies taken by all the other players except $i$ (hence the $_{-i}$)

In a game, each player $i$ chooses a strategy $s_i \in S_i$ in such a way that the player's utility function $u_i(s, \vec{s}_{-i})$ is optimized.

For simplicity, we focus on static games, meaning every game is considered as a one-shot. We also consider non-cooperative games, meaning there every player makes its decisions independent, and is only interested in its own benefits.

## 4.2  Definition 2: a Nash Equilibrium

As every user tries to optimize his utility function, there might be an impact on the utility function of other users, possibly causing a chain of events. In the worst case, this chain of events is infinite. It is highly desirable that in a limited number of steps a state is reached in which all users have optimized their utility function and decide not to take any further actions. This state is called a Nash Equilibrium and is defined as follows:

A Nash equilibrium of a game is a vector of strategies $\vec{s}^* \in S$ such that for every player $i$ holds:

$$u_i(s_i^*, \vec{s}_{-i}^*) \geq u_i(s_i, \vec{s}_{-i}^*), \forall s_i \in S_i \tag{1}$$

This is a very important concept in solving game theoretical problems. Indeed, a Nash equilibrium is a stable situation, in which no user unilaterally can improve its utility, given that everybody else's actions are fixed. There are some challenges as well: First of all, not every game has a Nash Equilibrium. Secondly, it is not guaranteed that the Nash equilibrium of a game is the 'desired' outcome of the game. In the following paragraph this is shown for a game called the Prisoner's dilemma, introducing the concept of a payoff table.

## 4.3  Payoff tables

An often quoted example of a game is called the Prisoner's dilemma: Two gang members are arrested for a big crime (sentence of 5 years in prison) and a small crime (sentence of 1 year in prison). There is enough evidence for the small crime, but not for the big crime. For the big crime to be sentenced, one gangster would have to betray the other. If both gangsters remain silent, the only sentence would be for the small crime. A police officer separately offers both gangsters freedom for betraying the other. If one of the gangsters talks, this can set him free. But if both gangsters talk, they can both be charged against the big crime, and will only receive a small sentence reduction.

This game has three possible outcomes:

1. Both gangsters remain silent and have 1 year prison
2. One gangster betrays the other: one person is free, the other gets 6 years of prison
3. Both gangsters betray each other: both gangsters get 5 years of prison.

Considering every gangster is only concerned about his own outcome, and considering the game is independent of any other events (no revenge actions later for example), we can make the following table with possible outcomes for the first gangster.

| **Years in prison, gangster 1** | | |
| --- | --- | --- |
|  | Action of gangster 2 | |
| Action of gangster 1 | Remain silent | Betray the other |
| Remain silent | 1 year | 6 years |
| Betray the other | 0 years | 5 years |

**Table 1: Prisoner's Dilemma - possible outcomes for gangster 1**

Gangster 1 only has impact on his own choice. He can either remain silent or he can betray the other. To make the right choice, he has to consider the possible actions of gangster 2, and decide on the best response. Every column is a possible action of gangster 2. If gangster 2 remains silent, gangster 1 can also remain silent and face 1 year of prison. Or he can betray the other, and get 0 years of prison. The latter is the best response, and is underlined in the table. The best response for the second column can be found the same way: If gangster 2 betrays gangster 1, gangster 1 can remain silent and get 6 years of prison, or he can betray the other, and get only 5 years of prison. Also in this case, betraying the other is the best option.

The same table can be made for gangster 2

| **Years in prison, gangster 2** | | |
|---|---|---|
| | Action of gangster 2 | |
| Action of gangster 1 | Remain silent | Betray the other |
| Remain silent | 1 year | 0 years |
| Betray the other | 6 years | 5 years |

Table 2: Prisoner's Dilemma - possible outcomes for gangster 2

Combining both tables makes it possible to find the Nash equilibrium. If both players consider the same vector of strategies as best response, a Nash equilibrium is found. Indeed, none of the players can improve its outcome by unilaterally changing his response.

| **Payoff table gangster 1 - gangster 2** | | |
|---|---|---|
| | Action of gangster 2 | |
| Action of gangster 1 | Remain silent | Betray the other |
| Remain silent | 1 year - 1 year | 6 years - 0 years |
| Betray the other | 0 years - 6 years | **5 years - 5 years** |

Table 3: Payoff table Prisoner's Dilemma with Nash Equilibrium in bold

The above definition describes a *pure*-strategy Nash equilibrium. *Mixed*-strategy Nash equilibria also exist, in this case multiple actions are involved, each with a certain probability. For simplicity, this paper only considers pure strategy Nash equilibria.

It can be considered positive for society to have both gangsters in prison, but from point of view of the game players (the gangsters) this is not the best situation. Intuitively it makes more sense if both players remained silent, but in that case, one player can improve its outcome by changing strategy.

When using game theory to solve demand side management challenges, Nash equilibria are searched for and considered the solution of the game. However, it is important to investigate that this equilibrium indeed has positive effects on the players (for example by lowering their energy cost) and on society (for example by lowering the peak-to-average ratio, diminishing the need for extra power plants).

# 5 Mathematical Model

It is assumed that (based on historical usage and on input from the household) every smart meter knows the household's energy requirements of the next day. Smart meters will try to schedule the shiftable appliances in such a way that the household's energy cost is minimized. This demand side management challenge is called the day-ahead load-scheduling problem. This section introduces a mathematical model based on game theory, representing this load-scheduling problem [5][6][7][8].

- The set of players $N$, from the definition of a game, is in this case the set of households.
- The utility function $(u_i)_{i \in N}$, is in this case the electricity cost of player $i$.
- The set of possible strategies $(S_i)_{i \in N}$ for a player $i$, is his energy usage profile throughout the day.

## 5.1 Strategies

If we divide the day into 24 periods of one hour, every possible action for every player would be a vector of 24 energy loads. This total energy load of a player is the sum of the energy loads of all appliances owned by the player.

Every player $i$ has a set of appliances $A_i$. For every appliance $a \in A_i$ a possible energy consumption scheduling vector is defined:

$$\vec{x}_{i,a} \triangleq [x_{i,a}^1, x_{i,a}^2, \ldots, x_{i,a}^{24}] \quad (2)$$

The total energy usage profile of player $i$ is its strategy $s_i$ in the game, and can be denoted as

$$\vec{s}_i \triangleq [s_i^1, s_i^2, \ldots, s_i^{24}] = \left[ \sum_{a \in A_i} x_{i,a}^1, \sum_{a \in A_i} x_{i,a}^2, \ldots \sum_{a \in A_i} x_{i,a}^{24} \right] = \sum_{a \in A_i} \vec{x}_{i,a} \quad (3)$$

Fictitious example with energy expressed in kWh:

- Laundry machine: 1kWh (shiftable load, duration 1 hour)
- Dryer: 1 kWh (shiftable load, duration 1 hour)
- Dish Washer: 1 kWh (shiftable load, duration 1 hour)
- Electric car: 20 kWh (shiftable load, maximum power level 4kW, duration 5 hours)
- Non shiftable loads (television, lights, computer, coffee machine, etc.): 6 kWh

In this example, consider the non shiftable loads as an energy consumption of 1 kWh every hour from 17h till 23h.

Let's choose the laundry machine to start at 16h, the dryer at 18h, the dishwasher at 21h and the electric car charges from midnight till 5h. All components are known to calculate the usage profile as in ( 3 ), it is the final row in Table 4

| Appliance | Load vector | | | | | | | | | | | | | | | | | | | | | | | |
|---|---|---|---|---|---|---|---|---|---|---|---|---|---|---|---|---|---|---|---|---|---|---|---|---|
| | 0 | 1 | 2 | 3 | 4 | 5 | 6 | 7 | 8 | 9 | 10 | 11 | 12 | 13 | 14 | 15 | 16 | 17 | 18 | 19 | 20 | 21 | 22 | 23 |
| Laundry machine | | | | | | | | | | | | | | | | | 1 | | | | | | | |
| Dryer | | | | | | | | | | | | | | | | | | | 1 | | | | | |
| Dish Washer | | | | | | | | | | | | | | | | | | | | | | 1 | | |
| Electrical car | 4 | 4 | 4 | 4 | 4 | | | | | | | | | | | | | | | | | | | |
| Not shiftable loads | | | | | | | | | | | | | | | | | | 1 | 1 | 1 | 1 | 1 | 1 | |
| Energy usage profile | 4 | 4 | 4 | 4 | 4 | 0 | 0 | 0 | 0 | 0 | 0 | 0 | 0 | 0 | 0 | 0 | 1 | 1 | 2 | 1 | 1 | 2 | 1 | 0 |

**Table 4: Example of energy scheduling strategy**

Another possible profile (so another strategy of player $i$) would be to start charging the car one hour later, and to use the dish washer at 18h and the dryer at 21h. This is shown in Table 5.

| Appliance | Load vector | | | | | | | | | | | | | | | | | | | | | | | |
|---|---|---|---|---|---|---|---|---|---|---|---|---|---|---|---|---|---|---|---|---|---|---|---|---|
| | 0 | 1 | 2 | 3 | 4 | 5 | 6 | 7 | 8 | 9 | 10 | 11 | 12 | 13 | 14 | 15 | 16 | 17 | 18 | 19 | 20 | 21 | 22 | 23 |
| Laundry machine | | | | | | | | | | | | | | | | | 1 | | | | | | | |
| Dryer | | | | | | | | | | | | | | | | | | | | | | 1 | | |
| Dish Washer | | | | | | | | | | | | | | | | | | | 1 | | | | | |
| Electrical car | | 4 | 4 | 4 | 4 | 4 | | | | | | | | | | | | | | | | | | |
| Not shiftable loads | | | | | | | | | | | | | | | | | | 1 | 1 | 1 | 1 | 1 | 1 | |
| Energy usage profile | 0 | 4 | 4 | 4 | 4 | 4 | 0 | 0 | 0 | 0 | 0 | 0 | 0 | 0 | 0 | 0 | 1 | 1 | 2 | 1 | 1 | 2 | 1 | 0 |

**Table 5: Alternative example of energy scheduling strategy**

A lot of different strategies are possible. 1 shiftable load has 24 possible start times. Adding a second shiftable load, means another 24 possible start times for this second load. Combined with all possible start times of the first load, this means there are 24*24 = 576 different scheduling possibilities for two shiftable loads. For the 4 shiftable loads in the above example, this means $24^4$ = 331.776 possibilities. Some strategies will end up in the same total energy usage profile. Even with part of the 331.776 possibilities ending up in the same energy usage profile, there will be thousands of possibilities, just for this one user.

In this game, the players are the households, and the possible strategies are the total load profiles, the sum in the table. As seen from the example, if the user decides to switch the timing of the dish washer with the timing of the dryer, the total energy usage profile remains unchanged. In this game setting, the outside world does not have to know exactly which appliance is requiring the energy, which is positive towards privacy.

Other games can be considered (for example in [5]) where every appliance plays the game individually, and the possible individual load profiles of the appliances are considered as strategies of the game. Each appliance tries to minimize its cost. This is an even more distributed approach.

### 5.2 Constraints

Until now every shiftable load could theoretically start at any time. In reality, this will not be the case. The car in the example will probably be occupied during the day, excluding all time slots between 9h and 17h as possible charging times. It also makes not much sense to start a dishwasher before eating dinner. The washing machine should be ready before 20h for example. The dryer cannot start before the laundry is done, so the dryer can start the earliest at 20h, and so on.

For every appliance, the player can set a minimum starting time and a maximum ending time. Between start and end, it should be possible for the appliance to finish its task. For example, one cannot start charging the car at 22h and expect it to be charged before 24h, because charging the car takes at least five hours. This information has to be added to the game formulation by means of constraints [6].

Consider $\alpha_{i,a}$ as the beginning and $\beta_{i,a}$ as the end of the time interval in which appliance $a \in A_i$ (set of appliances of player $i$) can be scheduled. Outside this interval, the energy usage of the appliance $a$ is 0 :

$$x_{i,a}^h = 0, \forall h \notin \{\alpha_{i,a}, \dots, \beta_{i,a}\} \tag{4}$$

While switched on, the energy usage of the appliance will be between a minimum (standby) power usage $\gamma_{i,a}^{min}$ and maximum power level $\gamma_{i,a}^{max}$. This can also be added as a constraint.

$$\gamma_{i,a}^{min} < x_{i,a}^h < \gamma_{i,a}^{max}, \qquad \forall h \in \{\alpha_{i,a}, \dots, \beta_{i,a}\} \tag{5}$$

Adding up the energy usage of every hour for appliance $a$ will result in the fixed total energy usage of appliance $a$, denoted as $E_{i,a}$.

$$\sum_{h=\alpha_{i,a}}^{\beta_{i,a}} x_{i,a}^h = E_{i,a} \tag{6}$$

For example for the electric car, the user could require the car to be charged between midnight and 8h, and a fully charged car could require 20 kWh of energy. If the maximum charging power is 4kW, and the standby power level is 0,01KW, this could be denoted as:

$$x_{car_i}^h = 0, \forall h \notin \{0, \dots, 8\}$$

$$0{,}01 < x_{car_i}^h < 4, \forall h \in \{0, \dots, 8\} \tag{7}$$

$$\sum_{h=0}^{8} x_{car_i}^h = 20$$

When charging a car, it is realistic to assume the energy usage per hour can be any value between 0,01kWh and 4kWh. For other appliances this constraint is a bit too loose [7]. If a dishwasher is started, it will run a few hours, and go over different phases of energy usage. For example, the second hour might require more heat and thus more energy. Once started, it cannot be paused. The required energy for the hours after start is known. For this kind of shiftable appliances a fixed load profile will have to be taken into account after the appliance has been started.

$$\vec{l}_{i,a} \triangleq [l_{i,a}^{start}, \dots, l_{i,a}^{end}] \tag{8}$$

In case of a dish washer that runs for three hours, this load profile could be

$$\vec{l}_{dishwasher_i} = [0.5, 1, 0.5] \tag{9}$$

Load profiles can also be put into constraints [5]. In this case, the load scheduling game is only about choosing the appliance start time. After start, the energy usage of the appliance is exactly its fixed load profile $\vec{l}_{i,a}$.

## 5.3 Utility Function

Every player $i$ tries to optimize his utility function $(u_i)_{i \in N}$. In this case the optimization is about looking for the minimum electricity cost.

One way to minimize electricity cost, is to use less electricity. This is a very good idea, but it is not what the game is about. This game is about scheduling electricity usage in such a way that the electricity cost is lower. This can be achieved by using electricity when it is as cheap as possible. If the price of electricity would be the same all day long, there would be no optimization needed, because the total electricity cost would always be the same.

Consider hourly electricity prices. The utility function $(u_i)_{i \in N}$ of every player $i$ is the electricity cost of this player. This cost is the sum of 24 hourly energy costs, being "energy price" times "energy used". The component "energy used" is known from ( 3 ) this is the strategy of the player:

$$\vec{s}_i \triangleq [s_i^1, s_i^2, \ldots, s_i^{24}] = [\sum_{a \in A_i} x_{i,a}^1, \sum_{a \in A_i} x_{i,a}^2, \ldots \sum_{a \in A_i} x_{i,a}^{24}] \triangleq \sum_{a \in A_i} \vec{x}_{i,a} \qquad (3), \text{repeated for clarity}$$

The component "energy price" is considered a price per hour, denoted as

$$\vec{p} \triangleq [p^1, p^2, \ldots, p^{24}] \qquad (10)$$

Multiplying ( 3 ) and ( 10 ) delivers the energy cost of player $i$

$$(u_i)_{i \in N} = \vec{p} \cdot \vec{s}_i \qquad (11)$$

$$= p^1 \cdot s_i^1 + p^2 \cdot s_i^2 + \cdots + p^{24} \cdot s_i^{24}$$

$$= p^1 \cdot \sum_{a \in A_i} x_{i,a}^1 + p^2 \cdot \sum_{a \in A_i} x_{i,a}^2 + \ldots + p^{24} \cdot \sum_{a \in A_i} x_{i,a}^{24}$$

$$= \sum_{h=1}^{24}(p^h \cdot \sum_{a \in A_i} x_{i,a}^h)$$

The price vector in ( 10 ) is the same for all players of the game, but it can change depending on the energy scheduling of all players. From section 2.2.2, we know that because of peak plants the energy production cost per unit energy is more expensive if there is a higher demand. The electricity production cost at a time h is thus a convex function of the total electricity used at time h. The production cost serves as a base for the electricity price, in such a way that the electricity price also becomes a convex function. This gives a motivation for shiftable appliances to move away from the peak. From a mathematical point this convexity is an advantage, because it turns out to be easier to find an optimum for convex functions. Convex programming techniques exist, and are often very helpful when looking for an optimum.

The price at time slot h is a function of the total load. The total load is the sum of all loads of all appliances, for all users, at time slot h, and is thus

$$p^h \triangleq f(\sum_{i \in N} \sum_{a \in A_i} x_{i,a}^h) \qquad (12)$$

An example of a quadratic price function could be ( 13 ) with $\alpha$ a constant.

$$f(y) = \alpha \cdot y^2 \qquad (13)$$

## 5.4 Formulation and solution of the game

The set of possible strategies $(S_i)_{i \in N}$ for a player $i$ taking into account the constraints follows immediately from ( 3 )( 4 )( 5 )( 6 ) and is denoted as

$$(S_i)_{i \in N} = \begin{cases} \vec{s}_i \triangleq [s_i^1, s_i^2, \dots, s_i^{24}] = \left[ \sum_{a \in A_i} x_{i,a}^1, \sum_{a \in A_i} x_{i,a}^2, \dots \sum_{a \in A_i} x_{i,a}^{24} \right] = \sum_{a \in A_i} \vec{x}_{i,a} \\ x_{i,a}^h = 0, \forall h \notin \{\alpha_{i,a}, \dots, \beta_{i,a}\} \\ \gamma_{i,a}^{min} < x_{i,a}^h < \gamma_{i,a}^{max}, \quad \forall h \in \{\alpha_{i,a}, \dots, \beta_{i,a}\} \\ \sum_{h=\alpha_{i,a}}^{\beta_{i,a}} x_{i,a}^h = E_{i,a} \end{cases} \quad (14)$$

The utility function $(u_i)_{i \in N}$, is the electricity cost of player $i$. The game $G$ is for each player to minimize this total energy cost. Substituting ( 12 ) into ( 11 ), the game can be formulated as:

$$G: \min_{\vec{s}_i} u_i(\vec{s}_i, \vec{s}_{-i}) = \sum_{h=1}^{24} \left( p^h \cdot \sum_{a \in A_i} x_{i,a}^h \right) = \sum_{h=1}^{24} \left( f\left( \sum_{i \in N} \sum_{a \in A_i} x_{i,a}^h \right) \cdot \sum_{a \in A_i} x_{i,a}^h \right), \forall i \in N \quad (15)$$

The most extensive way of solving this game, would be to list each possible load profile for every player and make a sum for every possible combination of these load profiles. This permits to calculate the hourly prices using ( 12 ), and thus also the total energy costs for each player for every combination of possible load profiles. This information can be entered into a payoff table to look for one or more Nash equilibria. This very cumbersome approach is demonstrated in paragraph 6.1. A faster way to solve the game is to use an algorithm that step by step evolves towards a Nash Equilibrium. Several algorithms and accompanying pitfalls are explained in paragraph 6.2.

# 6 Examples and Results

## 6.1 Simple example with Payoff Table

The game of section 5.4 is solved for the following scenario with 2 players and 3 timeslots [7].

Player 1, denoted as $pl1$ has the following loads:

- A fixed load of 1kWh at t = 1
- A shiftable load of 2kWh

Player 2, denoted as $pl2$ has the following loads:

- A fixed load of 5kWh at t = 2
- A shiftable load of 2,5kWh

Players $pl1$ and $pl2$ both have three possible strategies: The shiftable load can be executed in timeslot 1, 2 or 3. This results in 3 x 3 = 9 possibilities to be investigated.

The energy price is a function based on the total energy load per timeslot, as in ( 12 ). In ( 16 ) a simple quadratic price function is chosen.

$$f(y) \triangleq y^2 \qquad (16)$$

If both users would schedule the shiftable load in timeslot 1, this would result in the following prices:

At t=1, a total load of (1 + 2 + 2,5) kWh = 5,5kWh results in a price of 30,25 c€/kWh

At t=2, a total load of 5kWh, resulting in a price of 25 c€/kWh

In the Table 6 the cost for every player is calculated:

| timeslot | pl1 fixed | shiftable | total | price | cost | pl2 fixed | shiftable | total | price | cost | total load | price |
|---|---|---|---|---|---|---|---|---|---|---|---|---|
| 1 | 1 kWh | 2 kWh | 3 kWh | 30,25 c€/kWh | € 0,91 | 0 kWh | 2,5 kWh | 2,5 kWh | 30,25 c€/kWh | € 0,76 | 5,5 kWh | 30,25 c€/kWh |
| 2 | 0 kWh | 0 kWh | 0 kWh | 25,00 c€/kWh | € 0,00 | 5 kWh | 0 kWh | 5 kWh | 25,00 c€/kWh | € 1,25 | 5 kWh | 25,00 c€/kWh |
| 3 | 0 kWh | 0 kWh | 0 kWh | | € 0,00 | 0 kWh | 0 kWh | 0 kWh | | € 0,00 | 0 kWh | |
| | total energy used | 3 kWh | total cost pl1 | € 0,91 | | total energy used | 8 kWh | total cost pl2 | € 2,01 | | | |

**Table 6: Load profile cost scenario with quadratic price function**

If player two would decide to shift his shiftable load towards timeslot three, this results in cheaper energy costs for both players.

| timeslot | pl1 fixed | shiftable | total | price | cost | pl2 fixed | shiftable | total | price | cost | total load | price |
|---|---|---|---|---|---|---|---|---|---|---|---|---|
| 1 | 1 kWh | 2 kWh | 3 kWh | 9, c€/kWh | € 0,27 | 0 kWh | 0 kWh | 0 kWh | 9, c€/kWh | € 0,00 | 3,0 kWh | 9, c€/kWh |
| 2 | 0 kWh | 0 kWh | 0 kWh | 25,00 c€/kWh | € 0,00 | 5 kWh | 0 kWh | 5 kWh | 25,00 c€/kWh | € 1,25 | 5 kWh | 25,00 c€/kWh |
| 3 | 0 kWh | 0 kWh | 0 kWh | 6,25 c€/kWh | € 0,00 | 0 kWh | **2,5 kWh** | 2,5 kWh | 6,25 c€/kWh | € 0,16 | 2,5 kWh | 6,25 c€/kWh |
| | total energy used | 3 kWh | total cost pl1 | € 0,27 | | total energy used | 8 kWh | total cost pl2 | € 1,41 | | | |

**Table 7: Calculating total cost for every player after update load schedule player 2**

In both cases, the players each use the same total amount of energy, but the costs are much lower if the players schedule their shiftable loads in an intelligent way.

The resulting costs can be calculated for all 9 possible shiftable load combinations and are shown in the following payoff table. Best responses are underlined, Nash equilibria are in bold.

payoff table in €

| | t p2 = 1 | t p2 = 2 | t p2 = 3 |
|---|---|---|---|
| t p1 = 1 | 0,91 - 2,01 | 0,27 - 4,22 | **0,27 - 1,41** |
| t p1 = 2 | 1,10 - 2,76 | 1,82 - 6,77 | 0,99 - 2,61 |
| t p1 = 3 | **0,20 - 1,56** | 0,09 - 4,22 | 0,42 - 1,76 |

**Table 8: Payoff table simple load-scheduling game**

This game has two Nash equilibria. If player 1 places his shiftable load at t = 1, best response of player 2 is to place its shiftable load at t=3, and none of the players have an incentive to change their strategy.

The same holds if the two players would switch time slots: if player 1 places his shiftable load at t=3 and player 2 would use t=1.

We notice that these Nash equilibria are both interesting for the users. Looking at the total cost of energy, it is never lower than in the Nash equilibria. Both equilibria are valid solutions; one of the two will have to be chosen as the desired scheduling. In this case, a possible approach is to choose the Nash Equilibrium with the less total energy cost.

The solution is also positive for society. It is beneficial if peaks are avoided, because this would avoid extra costs and externalities from building extra peak power plants. An indication on how well peaks are avoided is the peak-to-average ratio (PAR), defined as the ratio of the maximum energy usage to the average energy usage. For a perfectly flat load profile the PAR would be 1. A high PAR reflects a high peak in energy usage. Table 9 shows the PAR for all possible load profiles. The two found Nash equilibria have the lowest PAR ratio. This was expected, as pricing was chosen to shift energy usage away from peaks.

| PAR | t p2 = 1 | t p2 = 2 | t p2 = 3 |
|---|---|---|---|
| t p1 = 1 | 1,57 | 2,14 | **1,43** |
| t p1 = 2 | 2,00 | 2,71 | 2,00 |
| t p1 = 3 | **1,43** | 2,14 | 1,43 |

**Table 9: Peak to average ratio for all possible simple load scheduling game combinations**

## 6.2   Algorithms for N players

In real life, choosing between Nash equilibria as in section 6.1 will not often occur, simply because in general the search for Nash equilibria will stop as soon as one is found. Using the previous way of work with payoff tables to solve the game for the set of $N$ players would require an $N$-dimensional payoff table. With thousands of possible strategies for every player, and $N$ players, a payoff table with $(1000's)^N$ different possibilities would have to be investigated. It is clear that as soon as more players and appliances are added, the problem becomes too big to be solved by the extensive method of creating a payoff table [7].

Instead, algorithms that (hopefully) converge step by step to a Nash equilibrium are used. As soon as an equilibrium is found, the algorithm stops. In this paper, we investigate several algorithms. High level, the studied algorithms can all be described as followed [3][5][6][7][8]:

1. Every players chooses a starting usage profile for all appliances, creating the 24-hours energy load vector as in ( 3 ), i.e. the strategy
2. This information is communicated to the grid. Based on the starting strategy of all players, the total load per hour is known, which delivers the price per hour as in ( 12 ). Exact choice of which function to use as a price function differs. In [6] the hourly cost for the electricity company is used instead of the price.
3. Information on the total load and in some cases on the cost/price function is communicated to a player. This player can be selected random or via a round robin principle. Based on the communicated information, the selected player tries to lower his total energy cost.
4. This new load profiles will be communicated towards the grid, and will imply new total loads per hour, and thus new hourly prices / costs.
5. Steps 3 and 4 are repeated until (hopefully) an equilibrium is reached

Though high level steps are similar, there are some important differences between the studied algorithms. We investigate the different approaches in detail, and discuss advantages and disadvantages.

### 6.2.1 Algorithm used in [5]

It is considered very positive if the algorithm guarantees to converge step by step towards an equilibrium. In [5], one tries to proof this based on the principles of Generalized Ordinal Potential Games. For Generalized Ordinal Potential Games, a potential function $\Phi: S \rightarrow \mathbb{R}$ exists that maps any possible combination of strategies of all players to a real number and for every player holds that

$$u_i(s_i, \vec{s}_{-i}) > u_i(s'_i, \vec{s}_{-i}) \implies \Phi(s_i, \vec{s}_{-i}) > \Phi(s'_i, \vec{s}_{-i}) \qquad (17)$$

At can be mathematically proven that these games admit to the Finite Improvement Property, which implies that a sequence of best response updates (as in the described algorithm) converge to a pure equilibrium.

As a potential function $\Phi$, the total energy cost of all players is taken. If this game would be a Generalized Ordinal Potential Game, this would mean that by improving one player's energy cost ($u_i(s_i, \vec{s}_{-i})$ in the definition) it is guaranteed that the total energy cost of all players ($\Phi(s_i, \vec{s}_{-i})$ in the definition) will also lower.

[5] states the game as formulized in ( 15 ) would be a Generalized Ordinal Potential Game for price functions of the form

$$f(y) = \alpha y^\beta, \alpha > 0, \beta \geq 1 \qquad (18)$$

under the condition that the shiftable load of one user is much smaller than the total energy load of the hour.

Unfortunately, this is not correct. This can be proven with a counter example. Considering the scenario depicted in Table 10, with price function $f(y) = 0{,}25 * y$

| player i | | | | | sum all players except player i | | | | | total | | |
|---|---|---|---|---|---|---|---|---|---|---|---|---|
| h | fixed | shiftabl | total | price | cost | fixed | shiftabl | total | price | cost | load | price | cost |
| 1 | 3 kWh | 2 kWh | 5 kWh | 25,25 c€/kWh | € 1,26 | 96 kWh | 0 kWh | 96 kWh | 25,25 c€/kWh | € 24,24 | 101 kWh | 25,25 c€/kWh | € 25,50 |
| 2 | 0 kWh | 0 kWh | 0 kWh | 25,00 c€/kWh | € 0,00 | 100 kWh | 0 kWh | 100 kWh | 25,00 c€/kWh | € 25,00 | 100 kWh | 25,00 c€/kWh | € 25,00 |
| | | | | total cost pl1 | € 1,26 | | | | total cost all players except player i | € 49,24 | | total cost all players | € 50,50 |

**Table 10: Load profile cost scenario with linear price function**

It is player $i$'s turn. He receives price information, and information on the total load of all players. In this simple example only two time slots exists. Player $i$ will investigate if it is beneficial for him to shift the shiftable load of 2kWh towards timeslot 2. The price in timeslot 2 is slightly cheaper, which gives a motivation to shift the load. On the other hand, shifting the load towards timeslot 2 will increase the electricity price in timeslot 2, because the total energy load in timeslot 2 becomes higher. However, and this is very important, this increase of electricity unit price in timeslot 2 is more than nullified by a decrease of the electricity price in timeslot 1 (as the total load in timeslot 1 becomes lower).

Player $i$ will shift the load, and gain 1 cent. Unfortunately, all other players will in total loose 2 cents. Comparing Table 10 and Table 11, it is clear the total energy cost for all players has risen 1 cent!

### Table 11: Calculating total cost for every player after load shift player i, total cost rises

| | player i | | | | | sum all players except player i | | | | | total | | |
|---|---|---|---|---|---|---|---|---|---|---|---|---|---|
| h | fixed | shiftabl | total | price | cost | fixed | shiftabl | total | price | cost | load | price | cost |
| 1 | 3 kWh | 0 kWh | 3 kWh | 24,75 c€/kWh | € 0,74 | 96 kWh | 0 kWh | 96 kWh | 24,75 c€/kWh | € 23,76 | 99 kWh | 24,75 c€/kWh | € 24,50 |
| 2 | 0 kWh | **2 kWh** | 2 kWh | 25,50 c€/kWh | € 0,51 | 100 kWh | 0 kWh | 100 kWh | 25,50 c€/kWh | € 25,50 | 102 kWh | 25,50 c€/kWh | € 26,01 |
| | | | | total cost pl1 | € 1,25 | | | | total cost all players except player i | € 49,26 | | total cost all players | € 50,51 |

An additional condition in [5] stated that the shiftable load of one user is much smaller than the total energy load of the hour; in this case 2 kWh is much smaller than the total hourly load of about 100 kWh. Even if this is not yet small enough, the same scenario could be rewritten for any small $\varepsilon$ and any price function $f(y) = \alpha y$, as shown in Table 12 and Table 13 shifting $2\varepsilon$ from timeslot 1 to timeslot 2 for player $i$.

| | player i | | | | | sum all players except player i | | | | | total | | |
|---|---|---|---|---|---|---|---|---|---|---|---|---|---|
| h | fixed | shiftabl | total | price | cost | fixed | shiftabl | total | price | cost | load | price | cost |
| 1 | x | 2ε | x+2ε | α*(y+ε) | (x+2ε)*α*(y+ε) | y-x-ε | 0 | y-x-ε | α*(y+ε) | (y-x-ε)*α*(y+ε) | y+ε | α*(y+ε) | α*(y+ε)^2 |
| 2 | 0 | 0 | 0 | α*y | 0 | y | 0 | y | α*y | α*(y)^2 | y | α*y | α*(y)^2 |
| | | | | total cost player i | sum | | | | | sum | | total | sum |

### Table 12: Load scheduling game with small shiftable load – before shifting load

| | player i | | | | | sum all players except player i | | | | | total | | |
|---|---|---|---|---|---|---|---|---|---|---|---|---|---|
| h | fixed | shiftabl | total | price | cost | fixed | shiftabl | total | price | cost | load | price | cost |
| 1 | x | 0 | x | α*(y-ε) | x*α*(y-ε) | y-x-ε | 0 | y-x-ε | α*(y-ε) | (y-x-ε)*α*(y+ε) | y-ε | α*(y-ε) | α*(y-ε)^2 |
| 2 | 0 | **2ε** | 2ε | α*(y+2ε) | 2ε*α*(y+2ε) | y | 0 | y | α*(y+2ε) | y*α*(y+2ε) | y | α*(y+2ε) | α*(y+2ε)^2 |
| | | | | total cost player i | sum | | | | | sum | | total | sum |

### Table 13: Load scheduling game with small shiftable load – after shifting load

Subtracting the total cost after shift from total cost before shift:

$$\begin{aligned}&\alpha[(y-\varepsilon)^2 + (y+2\varepsilon)^2 - (y+\varepsilon)^2 - y^2]\\ &= \alpha[y^2 - 2y\varepsilon + \varepsilon^2 + y^2 + 4y\varepsilon + 4\varepsilon^2 - y^2 - 2y\varepsilon - \varepsilon^2 - y^2]\\ &= \alpha 4\varepsilon^2\end{aligned} \qquad (19)$$

Given that $\alpha > 0$ this implies $\alpha 4\varepsilon^2 > 0$ , so the total cost will always rise.

Subtracting the total cost of player $i$ after shift from total cost before shift:

$$\begin{aligned}&\alpha[x(y-\varepsilon) + 2\varepsilon(y+2\varepsilon) - (x+2\varepsilon)(y+\varepsilon)]\\ &=\alpha(xy - x\varepsilon + 2y\varepsilon + 4\varepsilon^2 - xy - x\varepsilon - 2y\varepsilon - 2\varepsilon^2)\\ &=\alpha(-2x\varepsilon + 2\varepsilon^2)\end{aligned} \qquad (20)$$

This term is <0 if $\varepsilon < x$

This means that, in the given set up, no matter how big the total hourly load y, as soon as the shiftable load $2\varepsilon$ of player $i$ is smaller than twice the fixed load $x$ of player $i$, it will beneficial for player $i$ to shift his shiftable load, although it will increase the total cost with $\alpha 4\varepsilon^2$.

It is beneficial to move shiftable loads away from the own fixed load profile. This is because shifting the load away will lower the total energy usage in the time slots with own fixed usage, and this lowers the player's energy cost in these time slots.

This could make the game go into an infinite loop, as shown in Table 14 and Table 15.

Table 14 shows the before situation – player 1 is on turn.

| | pl1 | | | | | pl2 | | | | | pl3 | | | | | total | |
|---|---|---|---|---|---|---|---|---|---|---|---|---|---|---|---|---|---|
| h | fixed | shiftable | total | price | cost | fixed | shiftable | total | price | cost | fixed | shiftable | total | price | cost | load | price |
| 1 | 6 kWh | 6 kWh | 12 kWh | 28,8 c€/kWh | € 3,46 | 0 kWh | 0 kWh | 0 kWh | 28,8 c€/kWh | € 0,00 | 0 kWh | 0 kWh | 0 kWh | 28,8 c€/kWh | € 0,00 | 12 kWh | 28,8 c€/kWh |
| 2 | 0 kWh | 0 kWh | 0 kWh | 28,8 c€/kWh | € 0,00 | 6 kWh | 6 kWh | 12 kWh | 28,8 c€/kWh | € 3,46 | 0 kWh | 0 kWh | 0 kWh | 28,8 c€/kWh | € 0,00 | 12 kWh | 28,8 c€/kWh |
| 3 | 0 kWh | 0 kWh | 0 kWh | 28,8 c€/kWh | € 0,00 | 0 kWh | 0 kWh | 0 kWh | 28,8 c€/kWh | € 0,00 | 6 kWh | 6 kWh | 12 kWh | 28,8 c€/kWh | € 3,46 | 12 kWh | 28,8 c€/kWh |
| | | | | total cost pl1 | € 3,46 | | | | total cost pl2 | € 3,46 | | | | total cost pl3 | € 3,46 | | |

**Table 14: Load scheduling game in infinite loop before shifting load**

Table 15 shows the after situation – player 1 has shifted 1 kWh to second timeslot.

| | pl1 | | | | | pl2 | | | | | pl3 | | | | | total | |
|---|---|---|---|---|---|---|---|---|---|---|---|---|---|---|---|---|---|
| h | fixed | shiftable | total | price | cost | fixed | shiftable | total | price | cost | fixed | shiftable | total | price | cost | load | price |
| 1 | 6 kWh | 5 kWh | 11 kWh | 24,2 c€/kWh | € 2,66 | 0 kWh | 0 kWh | 0 kWh | 24,2 c€/kWh | € 0,00 | 0 kWh | 0 kWh | 0 kWh | 24,2 c€/kWh | € 0,00 | 11 kWh | 24,2 c€/kWh |
| 2 | 0 kWh | **1 kWh** | 1 kWh | 33,8 c€/kWh | € 0,34 | 6 kWh | 6 kWh | 12 kWh | 33,8 c€/kWh | € 4,06 | 0 kWh | 0 kWh | 0 kWh | 33,8 c€/kWh | € 0,00 | 13 kWh | 33,8 c€/kWh |
| 3 | 0 kWh | 0 kWh | 0 kWh | 28,8 c€/kWh | € 0,00 | 0 kWh | 0 kWh | 0 kWh | 28,8 c€/kWh | € 0,00 | 6 kWh | 6 kWh | 12 kWh | 28,8 c€/kWh | € 3,46 | 12 kWh | 28,8 c€/kWh |
| | | | | total cost pl1 | € 3,00 | | | | total cost pl2 | € 4,06 | | | | total cost pl3 | € 3,46 | | |

**Table 15: Load scheduling game in infinite loop after shifting load**

Player 1 will start with shifting away 1 kWh from his own fixed usage. This lowers his own total energy cost with 46c€, but makes the total energy cost of all users rise with 14c€. This action of player 1 will trigger player 2 or 3 to do the same. This could start an infinite loop of shifting loads around, not converging to an equilibrium.

These counterexamples show that the game is not a Generalized Ordinal Potential Game and that convergence is not always guaranteed. As a workaround, an extra check could be included in the algorithm: if the new load profile of a player worsens the total electricity cost, this new load profile is not accepted, and another player is selected (or the current player gets a second try). However, this could cause another problem. Players could cheat, and lie about which loads are shiftable (see also section 6.4). If player 1 is not allowed to have a shiftable load of 1kWh in timeslot 2, he will state this load of 1 kWh is fixed.

Luckily, simulations in [5] based on realistic smart grid scenarios from real data of 100 Italian households do converge towards an equilibrium. It is shown that only a few iterations are needed to achieve an equilibrium. The average energy cost lowered with 22%, also the Peak-to-Average ratio was reduced with 22%.

[5] adds another interesting idea. Until now, every household was considered a player of the game. In [5] an additional approach is investigated, where every appliance is a player on its own. This game is less complex. Instead of trying to combine all household appliances to the perfect load profile for the household, every appliance takes its own decisions. Simulation results of this approach show the final electricity bill is a bit higher (about 3% higher than the approach where the households are players) but that calculation times are much lower (about ten times lower).

### 6.2.2   Algorithm used in [6] and [7]

This algorithm focuses on the total production cost for the electricity company, rather than on the cost for every user. This total production cost will be divided pro rata amongst all users, so every user will be charged the same electricity unit price, no matter if his electricity was in an expensive peak period or not. The idea is that lowering the total electricity production cost will also lower a player's fixed fraction of this cost, and thus give him a motivation to participate as much and honest as possible in the demand side management load shifting game.

The utility function for this game is not hourly prices times hourly load as in ( 11 ), the unit price is the same for all periods. This utility function is: "pro rata part" times "total cost" (times a fixed profit margin, left out for simplicity).

$$G: \min_{\vec{s}_i} u_i(\vec{s}_i, \vec{s}_{-i}) = \frac{\sum_{h=1}^{24}\sum_{a \in A_i} x_{i,a}^h}{\sum_{i \in N}\sum_{h=1}^{24}\sum_{a \in A_i} x_{i,a}^h} \sum_{h=1}^{24} C^h =, \forall i \in N \qquad (21)$$

In ( 21 ) the pro rata term $\frac{\sum_{h=1}^{24}\sum_{a \in A_i} x_{i,a}^h}{\sum_{i \in N}\sum_{h=1}^{24}\sum_{a \in A_i} x_{i,a}^h}$ is fixed, so only focus is on $\sum_{h=1}^{24} C^h$.

The production cost for the electricity company at hour $h$, denoted $c^h$, depends on the total load at hour $h$.

$$c^h \triangleq f(\sum_{i \in N} \sum_{a \in A_i} x_{i,a}^h) \qquad (22)$$

By shifting the energy use of shiftable appliances away from peaks, player $i$ has impact on this total load at hour $h$ and thus on the energy production cost at hour $h$. From section 2.2 we know that – because of use of expensive peak plants – a linear increase in electricity load will have a more than linear increase in total electricity production cost. This means that $c^h$ is a convex function, which is considered a big advantage according to [6]. It is stated that the convexity of $c^h$ mathematically implies that always one unique Nash equilibrium exists. Unfortunately this uniqueness of the Nash equilibrium is proven wrong by a counterexample in [7]. As discussed before, multiple Nash equilibria often means that only one of the will have the lowest total cost. Other Nash equilibria will be slightly inferior solutions.

On the bright side, the theory of Generalized Ordinal Potential Games from section 6.2.1 is now really valid. The only way to reduce the energy cost of player $i$ (being a fixed fraction of the total cost), is to reduce the total energy production cost itself, so ( 17 ) holds for this algorithm

with Φ being the total cost function

$$\Phi \triangleq \sum_{h=1}^{24} C^h \qquad (23)$$

The Finite Improvement Property now which implies that a sequence of best response updates (as in the described algorithm of 6.2) converges to a pure equilibrium. In every step of the algorithm, the energy production cost becomes lower or stays the same. This will converge in a limited number of steps towards an equilibrium.

A disadvantage of this approach is that everybody pays the same electricity unit price over all periods. If player $i$ wants to use expensive peak power, his electricity bill will slightly go up, together with everybody else's electricity bill. While the benefits of being able to use the peak power are purely for player $i$, the externalities of this choice are for all players.

Table 16 compares the results with a constant unit price and the results from section 6.1, an approach with hourly unit prices. Notice both algorithms have two Nash equilibria, showing the non-uniqueness of Nash equilibria by a counterexample. Comparing every field in the two tables, we notice the total cost is the same. Focusing on the two Nash equilibria, we notice the cost for player 1 is higher with constant unit prices. Player 1 has to pay part of the high production cost caused by the peak usage from player 2.

payoff table constant unit price (6.2.2)

|  | t p2 = 1 | t p2 = 2 | t p2 = 3 |
|---|---|---|---|
| t p1 = 1 | 0,83 - 2,08 | 1,28 - 3,21 | **0,48 - 1,20** |
| t p1 = 2 | 1,10 - 2,76 | 2,45 - 6,13 | 1,03 - 2,57 |
| t p1 = 3 | **0,50 - 1,26** | 1,23 - 3,08 | 0,62 - 1,55 |

payoff table hourly unit prices (6.1)

|  | t p2 = 1 | t p2 = 2 | t p2 = 3 |
|---|---|---|---|
| t p1 = 1 | 0,91 - 2,01 | 0,27 - 4,22 | **0,27 - 1,41** |
| t p1 = 2 | 1,10 - 2,76 | 1,82 - 6,77 | 0,99 - 2,61 |
| t p1 = 3 | **0,20 - 1,56** | 0,09 - 4,22 | 0,42 - 1,76 |

**Table 16: Comparing two payoff tables with different price functions**

Simulation results on 10 households with each 10 to 20 non-shiftable and 10 to 20 shiftable appliances showed an energy cost reduction of 18%. The peak-to-average ratio lowered with 17%. The algorithms converges after about 22 iterations, this is 2 iterations per user.

### 6.2.3 Algorithm used in [8]

This time an extra component is added to quantify the player's (in)tolerance towards the shifting of energy consumption. The utility function becomes

$$u_i(\vec{s}_i, \vec{s}_{-i}) = Value\ of\ Energy - Cost\ of\ Energy \quad (24)$$

The term "Cost of Energy" is the same as in ( 15 )

The term "Value of Energy" is defined as

$$V(X_i) \triangleq pE_{max,i}(1 - e^{-\omega_i E_i}) \quad (25)$$

With $p$ the average energy price, $E_{max,i}$ the maximum amount of energy player $i$ can consume, $E_i$ the total energy consumption of player $i$ in the game and $\omega_i$ the tolerance factor of player $i$ towards energy shifting. The higher $\omega_i$, the higher $V(X_i)$.

Although the intentions of including a valuation of energy are good, this component will not change the outcome of the game. The "Value of Energy" for a player $i$ is formula of parameters that remain constant during the game. The "Value of Energy" itself is thus also nothing more than a constant that is added to the game of ( 15 ) (with inverted sign). In mathematics, optimizing $f(x)$ or optimizing $f(x) + constant$ delivers exactly the same results.

The following function is used as pricing function in ( 12 ):

$$f(y) = \alpha.y.\log(y + 1) \quad (26)$$

This is a comparable function as in Section 6.2.1. The results are also comparable. Simulations are run for 5 to 500 households. Every household has 10 to 15 shiftable and 10 to 15 non-shiftable appliances. The peak-to-average ratio is lowered with about 27%. The energy cost reduction is not mentioned.

### 6.3 Distributed approach

All studied algorithms use a distributed approach. A central unit in the grid will select the next player and communicate total load and price information, but the players calculate the energy usage profile themselves.

This approach has big advantages.

First of all every player will choose his strategy taking into account his own constraints. Other players do not need to know about these constraints or not even about the strategy. Every player has the freedom to add additional personal constraints. This is a big advantage over a centralized approach, where all players would have to communicate all possible variables and constraints to a central unit that does the calculations.

In most cases, only the total load of all other users is communicated towards a player. This is an advantage on privacy. Every player does not need to know the exact details on the energy usage of a particular other player.

## 6.4 Cheating

When formulating a game, it is important to make it lying proof. In [7] it is correctly stated that if users would prefer running a certain shiftable appliance a fixed time, they could simply lie and denote it as a non-shiftable load. This is not really fair. Other users will then be bullied away from this timeframe, and always encounter small inconveniences while liars would never have to wait.

A possibility to diminish cheating would be to extend the formulation of the game. Instead of only looking at the cost, one could also introduce formulas that put a value to running a certain appliance in a certain timeframe. In that case it becomes more interesting to state a washing machine as shiftable, but with a high preference to run it during the day, than lying and putting the load profile of the washing machine as non-shiftable.

However, note that this introduces other challenges. Like is it socially acceptable that more affluent customers would always get the best timeframes? And while with the current game formulation the peak-to-average ratio of the electricity use goes down, with a changed game setting it might actually go up. It is important that the valuation of timeslot preferences is not too cheaply priced.

# 7 Future

## 7.1 Supply following appliances

A possible way to produce sustainable electricity is distributed generation of wind and solar power. This introduces new challenges. In the investigated studies, focus is on shaving off peaks and reducing the need for peak power plants. However, if a substantial part of the electricity comes from intermittent renewable power plants, additional focus should be on supply following approaches. It is not possible to schedule wind and solar energy production, and electricity still has to be consumed at the moment when it is produced. If in a time slot a lot of energy is supplied because of high wind and solar availability, shiftable appliances should follow.

Based on weather patterns, the available wind and solar power can be predicted. If this prediction is taken into account in the game, a similar approach as above could be applied. Only this time, the price function will not necessarily be convex, and it can become harder to find a Nash equilibrium.

## 7.2  Electricity Prosumers

More and more households not only consume energy, but (for example with solar panels) also produce energy. This distributed production of renewable energy adds a lot of extra possibilities to the game. A player $i$ would not only have to decide when to schedule his appliances, but also whether to use his own energy or the energy from the grid. He might even be able to sell his own produced energy to the grid. In [9] an algorithm is introduced that includes households with dispatchable and non-dispatchable energy sources, but this is left out of the (limited) scope of this paper.

If the number of variables becomes really large, the game could become too difficult to solve. In this case, a possible way of work is to only schedule the large loads [7]. This game will be less complex. After this first step, smaller shiftable loads can be used to make the total energy load profile smoother.

## 7.3  Batteries and Electrical Vehicles

Until now in our models electricity had to be consumed by a household appliance at the moment when it was produced. It is possible to avoid this, by introducing storage. The players of the game could try to store energy when it is cheap, and use stored energy when energy on the market is too expensive. This adds again a new dimension to the game.

Deciding on whether to buy storage (and how much) becomes a game on its own [10]. Imagine everybody else has storage. This would mean as soon as the slightest rise in electricity price is expected, somebody will be able to use his storage to shift electricity usage away from the peak. In fact the electricity usage (and thus the electricity price) would become constant. In this scenario with flat electricity prices caused by lots of households with storage, one single household without storage would not have impact on the flat price. This could cause a chain of reactions with people deciding to save money by not buying storage. In [10] it is calculated that a scenario with 38% of households owning storage is the most beneficial for everybody.

Electrical vehicles could become extremely useful as energy storage. The amount of energy an electrical vehicle can store is around the total daily energy usage of an average household. If it would be possible to use this energy when appropriate (for example when energy from the grid becomes too expensive), this could be an important cost saving. One step further would be to allow selling the energy stored in batteries to the grid, a similar scenario as in 7.2.

## 7.4  Transition Period

Today, most houses in Belgium do not have a smart meter. Also, most household appliances are not able to communicate with a smart meter in order to send expected usage, or in order to be scheduled. An average household appliance has a lifetime of ten up to fifteen years. It will take some time before we live in a smart house connected to a smart grid. Taking these inevitable needs into account in production of new appliances and in construction of new houses, would be a great first step.

# 8   Conclusions

It is expected that in the future, electricity companies will start charging a fluctuating electricity price that is related to the real production cost [1]. This is a form of demand side management, because as soon as these dynamic pricing schemes are applied, users will try to benefit from the cheapest prices and avoid peak prices, supported by current possibilities in electronics and ICT. The outcomes of this future set up were investigated in this paper using game theoretic concepts to solve the day-ahead load-scheduling problem.

A complete elaboration, by means of a payoff table, was done for a simple example. It is shown that for more complicated examples, this extensive way of work is too complex. As an alternative, best response algorithms are introduced. Based on the literature, several algorithms are investigated.

Both theoretical and practical results are critically challenged:

In [5] a dynamic price function of the form $f(y) = \alpha y^\beta$ is introduced, and a convergence towards an equilibrium is guaranteed. By means of counterexamples, we show convergence is not always the case.

In [6] every user pays a pro rata part of the production cost (possibly multiplied by a profit margin). Convergence towards a unique equilibrium is guaranteed. Inspired by [7] this paper repeats that, although convergence towards an equilibrium is correct, this equilibrium is not unique and the algorithm could converge to a less-perfect solution. Moreover, we show that an undesired side effect of this algorithm is that the cost of peak usage is distributed amongst all players. While the benefits of being able to use the peak power are purely for one player, the externalities of this choice are for all players. This is not fair.

[8] introduces the concept of "Value of Energy". Although this is a very good idea, we show that the suggested formula for "Value of Energy" will have no impact on the end result. We also warn that introducing preference for certain time frames makes the problem more complex, and could result in worse instead of better peak-to-average ratios.

On the bright side, literature shows that simulations of all investigated algorithms do converge to an equilibrium, and that the energy cost and peak-to-average ratio is reduced with around 20%.

However, the remarks made in this paper show that it is still a huge challenge to model a game that is fair and that allows to put a value on energy usage preferences, together with an accompanying algorithm that is guaranteed to converge to a (unique) equilibrium, preferable with the lowest possible peak-to-average ratio and lowest possible energy cost.

# 9 List of figures and tables